\documentclass[a4paper,11pt]{article}
\usepackage{pos}

\usepackage{subfigure}

\newcommand{\pos}[1]{\href{https://pos.sissa.it/contribution?id=#1}{\ttfamily #1}} 

\title{Another look at the three-gluon vertex in the minimal Landau gauge}
 \ShortTitle{Landau gauge three-gluon vertex}

 \author[a,b]{Guilherme T. R. Catumba}
 \author[b]{Orlando Oliveira}
\author*[b]{Paulo J. Silva}

\affiliation[a]{IFIC - University of Valencia - Spain}

\affiliation[b]{CFisUC, Department of Physics, University of Coimbra, 3004-516 Coimbra, Portugal}
\emailAdd{gtelo@ific.uv.es}
\emailAdd{orlando@uc.pt}
\emailAdd{psilva@uc.pt}

\abstract{The lattice three-gluon vertex in the Landau gauge is revisited using a large physical volume $\sim(8\textrm{fm})^4$ and a large statistical ensemble. The improved calculation explores the symmetries of the hypercubic lattice to reduce the statistical uncertainties and addresses the evaluation of the lattice artefacts. Special attention is given to the low energy behaviour of the vertex and its relation to ghost dominance.}

\FullConference{%
 The 38th International Symposium on Lattice Field Theory, LATTICE2021
  26th-30th July, 2021
  Zoom/Gather@Massachusetts Institute of Technology
}

\begin{document}
\maketitle

\section{Introduction}

The three-gluon vertex is one of the QCD fundamental Green's functions. This vertex allows the computation of the strong coupling constant and the measurement of a static potential between color charges. Herein we report on an upgrade of the lattice computation of this vertex performed by some of the authors in \cite{duarte2016, proc2016}.

The  three-gluon correlation function $G^{a_1 a_2 a_3}_{\mu_1 \mu_2 \mu_3} (p_1, p_2, p_3)$ is given by
\begin{equation}
   \langle A^{a_1}_{\mu_1} (p_1) \, A^{a_2}_{\mu_2} (p_2) \, A^{a_3}_{\mu_3} (p_3) \rangle =   V \, \delta( p_1 + p_2 + p_3) ~
   {G^{a_1 a_2 a_3}_{\mu_1 \mu_2 \mu_3} (p_1, p_2, p_3)}
\end{equation}
and can be written in terms of the gluon propagator $D^{ab}_{\mu\nu}(p^2)$ and the one-particle irreducible (1PI) vertex $\Gamma$ using
\begin{equation}
  {G^{a_1a_2a_3}_{\mu_1\mu_2\mu_3} (p_1, p_2, p_3)}  =   D^{a_1b_1}_{\mu_1\nu_1}(p_1) ~ D^{a_2b_2}_{\mu_2\nu_2}(p_2) ~ D^{a_3b_3}_{\mu_3\nu_3}(p_3) 
    {\Gamma^{b_1b_2b_3}_{\nu_1\nu_2\nu_3} (p_1, p_2, p_3)} .
\end{equation}
Bose symmetry requires the 1PI vertex to be symmetric under permutations of any pair $(p_i, a_i, \mu_i)$. Given that
\begin{equation}
 \Gamma^{a_1 a_2 a_3}_{\mu_1 \mu_2 \mu_3} (p_1,  p_2, p_3) = f_{a_1 a_2 a_3} \Gamma_{\mu_1 \mu_2 \mu_3} (p_1, p_2, p_3)
\end{equation}
then the function $\Gamma_{\mu_1 \mu_2 \mu_3} (p_1, p_2, p_3)$ must be antisymmetric under the interchange of any pair $(p_i, \mu_i)$.

A complete description of $\Gamma_{\mu_1 \mu_2 \mu_3} (p_1, p_2, p_3)$ in the continuum requires six Lorentz invariant form factors, two associated to the transverse component $\Gamma^{(t)}$ 
and the remaining associated to the longitudinal $\Gamma^{(l)}$ \cite{ballchiu}.

\section{Asymmetric momentum configuration}

In this work we consider the computation of the three-gluon vertex in the asymmetric momentum configuration $p_2=0$, as in \cite{alles, duarte2016}. In this case, the correlation function can be written as
\begin{equation}
   G_{\mu_1\mu_2\mu_3} (p, 0, -p)  =    V \frac{N_c(N^2_c-1)}{4}  \left[D(p^2)\right]^2 \, D(0) \frac{\Gamma (p^2)}{3} ~ ~ p_{\mu_2} ~T_{\mu_1\mu_3} (p).
\end{equation}
The contraction of the Lorentz  $\mu_1$ and $\mu_3$ indices, together with the contraction with the momentum $p_\alpha$, gives
\begin{equation}
    G_{\mu \, \alpha \,\mu} (p, 0, -p) \, p_\alpha = V \frac{N_c(N^2_c-1)}{4}
   \, \left[D(p^2)\right]^2 \, D(0) ~~\Gamma (p^2) ~~ p^2 .
\end{equation}
From this expression it is possible to extract the form factor $\Gamma (p^2)$. However, a lattice measurement of  $\Gamma (p^2)$ requires the  computation of the ratio
\begin{equation}
 G_{\mu \alpha \mu} (p, 0, -p) p_\alpha  / \left[D(p^2)\right]^2 \, D(0) 
\end{equation}
and the extraction of $\Gamma (p^2)$ from this ratio will originate large statistical fluctuations at high momenta, where $D(p^2)$ becomes quite small. In fact, assuming Gaussian error propagation, it is possible to show that the statistical error on $\Gamma (p^2)$ behaves as $\Delta \Gamma(p^2) \sim p^4$ in the UV regime \cite{duarte2016}.

\section{Handling of noise, lattice artefacts}

In order to try to deal with the large statistical fluctuations at high momenta, we considered a few strategies \cite{guitese}:

\begin{itemize}
\item explore the ambiguity on the scale setting and perform a binning in the momentum --- all data points in each bin are replaced by a weighted average of the data points;
\item perform a $H(4)$ extrapolation of the lattice data \cite{becirevic1999, soto2009} --- such procedure is based on the  remnant $H(4)$ symmetry group
associated with a hypercubic lattice. On the lattice, a scalar quantity $F$ is a function of the $H(4)$ invariants
\begin{displaymath}
  p^2 = p^{[2]} = \sum_\mu p^2_\mu ,  \quad
  p^{[4]} = \sum_\mu p^4_\mu ,  \quad
  p^{[6]} = \sum_\mu p^6_\mu ,  \quad
  p^{[8]} = \sum_\mu p^8_\mu ,
\end{displaymath}
i.e. $F_{Lat} = F(p^{[2]}, p^{[4]}, p^{[6]}, p^{[8]})$. The continuum limit  will be given by $F(p^{[2]}, 0, 0, 0)$ up to corrections $\mathcal{O}(a^2)$. Having several data points for the same $p^2$ but different $p^{[4]}$, $p^{[6]}$, $p^{[8]}$, an extrapolation of $F_{Lat}$ to the continuum limit can be done, assuming that it can be written as a power series of the H(4) invariants. Note that, in this work, only a linear extrapolation in $p^{[4]}$ is considered.

\end{itemize}

\section{Lattice setup}

In this work we consider the $64^4$ ensemble of 2000 configurations already studied in \cite{duarte2016}, together with a $80^4$ ensemble of 1800 configurations, both generated with the Wilson gauge action at $\beta=6.0$. The rotation to the Landau gauge has been performed using the Fourier accelerated Steepest Descent method \cite{davies} implemented with the help of Chroma \cite{chroma} and PFFT \cite{pfft} libraries. The gluon field is computed using the definition
\begin{equation}
a g_0 A_\mu (x + a \hat{e}_\mu)  = \frac{ U_\mu (x) - U^\dagger (x)}{ 2 i g_0} 
      - \frac{\mbox{Tr} \left[ U_\mu (x) - U^\dagger (x) \right]}{6 i g_0} 
\end{equation}
with the momentum space gluon field given by
\begin{equation}
A_\mu (\hat{p}) = \sum_x e^{- i \hat{p} (x + a \hat{e}_\mu) } \, A_\mu (x + a \hat{e}_\mu) \,\,,\,\, \hat{p}_\mu = \frac{2 \, \pi \, n_\mu}{a \, L_\mu}.
\end{equation}

\section{Results}

\begin{figure}[h]
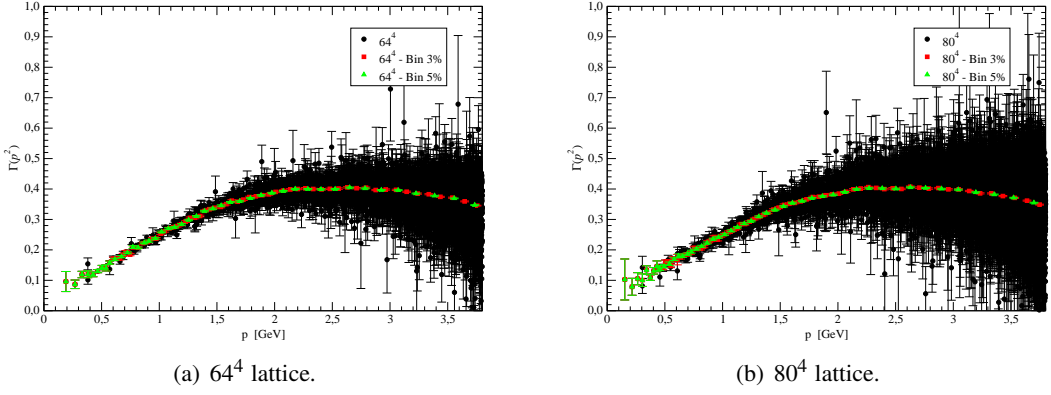
 
\vspace{0.55cm}
   \centering
   \subfigure[$64^4$ lattice.]{ \includegraphics[width=0.42\textwidth]{plots/gamma_64x4.eps} \label{binn64}} \qquad
   \subfigure[$80^4$ lattice.]{ \includegraphics[width=0.42\textwidth]{plots/gamma_80x4.eps} \label{binn80}}
  \caption{Original and binned data for $\Gamma (p^2)$.}
   \label{binned}
\end{figure}

In Figure \ref{binned} we compare the original and binned data for $\Gamma (p^2)$. The binning of the data suppresses the large statistical errors in the high momentum region and produces a well defined and smooth curve.

\begin{figure}[h] 
\vspace{0.65cm}
\begin{center}
\includegraphics[width=0.6\textwidth]{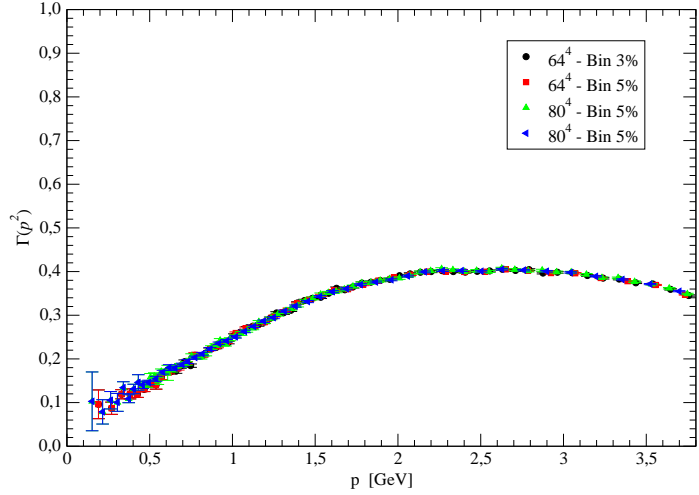}
\end{center}
 \caption{Comparison of binned data for  $\Gamma (p^2)$.}
   \label{binnedoverp2}
\end{figure}

Next, in Figure \ref{binnedoverp2} we compare the binned data for both lattices. The results of the two volumes agree within errors, suggesting that finite volume effects are small. 

\begin{figure}[h] 
\vspace{0.65cm}
\begin{center}
\includegraphics[width=0.6\textwidth]{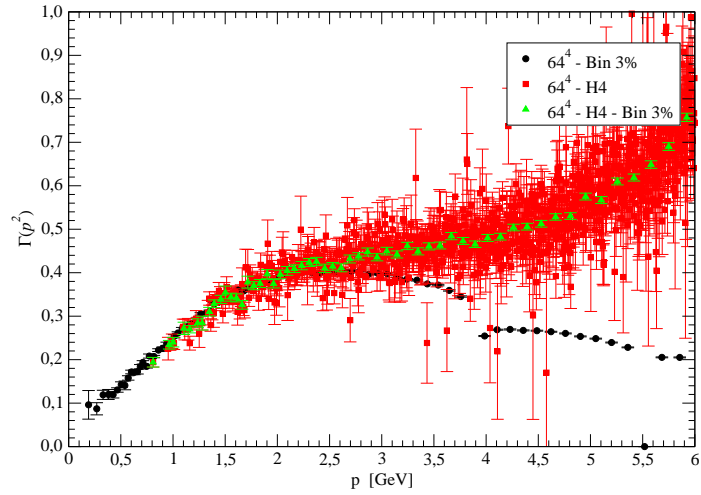}
\end{center}
 \caption{Results of the H(4) extrapolation of  $\Gamma (p^2)$ on the $64^4$ lattice volume.}
   \label{H4extr}
\end{figure}

\begin{figure}[h]
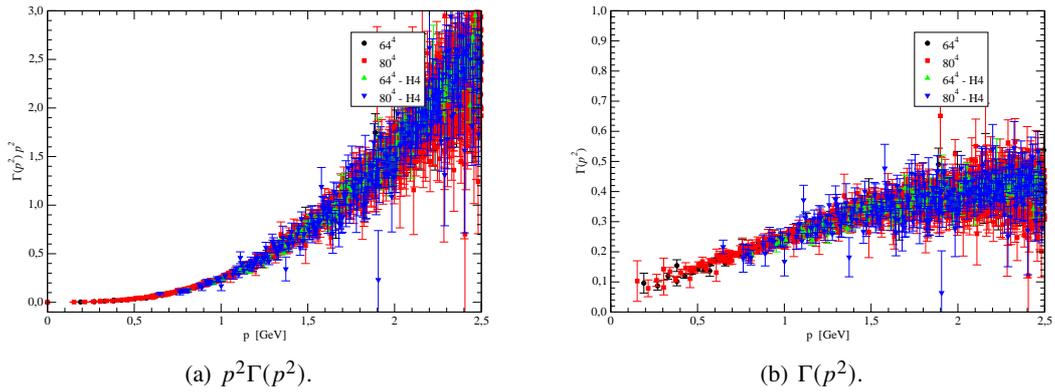
 
\vspace{0.55cm}
   \centering
   \subfigure[$p^2 \Gamma(p^2)$.]{ \includegraphics[width=0.42\textwidth]{plots/all_gamma.eps} \label{H4infra-p2G}} \qquad
   \subfigure[$\Gamma(p^2)$.]{ \includegraphics[width=0.42\textwidth]{plots/all_gamma_over_p2.eps} \label{H4infra-G}}
  \caption{Original and H(4) data for both lattice volumes for low momenta.}
   \label{H4infra}
\end{figure}

In Figure \ref{H4extr} we compare the H(4) extrapolation of the $64^4$ lattice data with the binning of the original data. We observe that the H(4) extrapolation pushes the vertex to higher values in the high momentum regime. Nevertheless, in the infrared region, the extrapolated data is compatible with the original data, for both lattice volumes --- see Figure \ref{H4infra}.

\section{Infrared behaviour of $\Gamma(p^2)$}

No zero crossing of $\Gamma(p^2)$, an indication of ghost dominance in the
infrared, is seen in the lattice data reported here. In order to check for
a change of sign in $\Gamma(p^2)$, in this  section we explore the infrared
behaviour of the lattice $\Gamma(p^2)$, using the $80^4$ data for momenta below 1GeV, and  fit the data to  $\Gamma_1(p^2)=A + Z \ln(p^2)$ and $ \Gamma_2(p^2)=A + Z \ln(p^2+m^2)$. The first one is a typical ansatz considered in recent studies to study the zero crossing, see \cite{guitese} for details, and the second one is a variant of the first one which includes an infrared logarithmic regularizing mass. 

\begin{figure}[h]
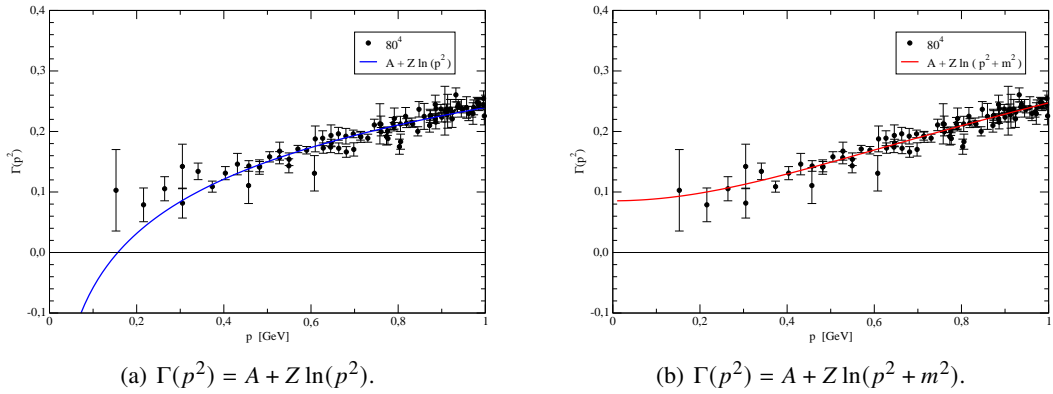
 
\vspace{0.55cm}
   \centering
   \subfigure[$\Gamma (p^2) = A + Z \ln(p^2)$.]{ \includegraphics[width=0.42\textwidth]{plots/gamma80-fit1.eps} \label{zerocrossing-fit1}} \qquad
   \subfigure[$\Gamma (p^2) = A + Z \ln(p^2+m^2)$.]{ \includegraphics[width=0.42\textwidth]{plots/gamma80-fit2.eps} \label{zerocrossing-fit2}}
  \caption{Infrared $80^4$ lattice data for $\Gamma(p^2)$ together with some fitting functions. }
   \label{zerocrossing}
\end{figure}

In Figure \ref{zerocrossing} we plot the best fits of the lattice data for both fitting functions, obtained through the minimization of $\chi^2/d.o.f.$ .
For $\Gamma_1(p^2)$, we got $\chi^2/d.o.f. = 1.23$ with $A=0.2395(16)$ and  $Z=0.0646(21)$. Accordingly the zero crossing occurs at  $p_o=157$MeV.
For $\Gamma_2(p^2)$, the parameters take the values $A=0.208(24)$, $Z=0.124(27)$ and $m=0.61(15)$GeV, with a  $\chi^2/d.o.f. = 0.95$. As shown in the right plot of Figure  \ref{zerocrossing}, in this case there is no zero crossing. 

\section{Conclusions and outlook}

In this paper we describe an improved calculation of the three gluon vertex on the lattice, for the asymmetric momentum configuration. We use two different lattice volumes  $(6.5$ fm$)^4$ and $(8.2$ fm$)^4$, with a common lattice  spacing of  $a = 0.102$ fm. We show that a H(4) extrapolation of the lattice data pushes the vertex to higher values in UV regime. We proceed with a  functional study in the infrared region, considering  some functional forms compatible with zero crossing and IR divergence.

Further momentum configurations will be explored in the near future.

\acknowledgments

This work was supported by national funds from FCT
Fundação para a Ciência e a Tecnologia, I. P., within the
Projects UIDB/04564/2020, UIDP/04564/2020, and CERN/FIS-COM/0029/2017.
G. T. R. C. acknowledges financial support from FCT
  under Project UIDB/04564/2020, and also  from the Generalitat Valenciana
  (genT program CIDEGENT/2019/040) and Ministerio de Ciencia e
  Innovacion PID2020-113644GB-I00.
P. J. S. acknowledges financial support from FCT
 under Contract CEECIND/00488/2017.
This work was granted access to the HPC resources of
the PDC Center for High Performance Computing at the
KTH Royal Institute of Technology, Sweden, made
available within the Distributed European Computing
Initiative by the PRACE-2IP, receiving funding from the
European Communitys Seventh Framework Programme
(FP7/2007-2013) under Grant agreement no. RI-283493.
The use of Lindgren has been provided under DECI-9
project COIMBRALATT. We acknowledge that the results
of this research have been achieved using the PRACE-3IP
project (FP7 RI312763) resource Sisu based in Finland at
CSC. The use of Sisu has been provided under DECI-12
project COIMBRALATT2. We acknowledge the
Laboratory for Advanced Computing at the University of
Coimbra \cite{lca} for providing access to the HPC resource
Navigator. The authors acknowledge Minho Advanced Computing Center
\cite{macc} for providing HPC resources that have contributed to
the research results reported within this paper. This work was
produced with the support of MACC and it was funded by FCT I.P.
under the Advanced Computing Project CPCA/A2/6816/2020, platform Bob.
This work was produced with the support of INCD \cite{incd} funded by FCT and
FEDER under the project 01/SAICT/2016 nº 022153.

\end{document}